\documentclass[11pt]{article}
\usepackage{moriond,epsfig}
\usepackage{amsmath,amssymb}
\usepackage{mathrsfs}
\usepackage{smaller}

\def\be{\begin{equation}}
\def\ee{\end{equation}}
\def\bea{\begin{eqnarray}}
\def\eea{\end{eqnarray}}

\DeclareMathAlphabet{\mathbbf}{OT1}{cmr}{bx}{it}

\begin{document}
\vspace*{4cm}
\title{BEAUTY PRODUCTION WITH THE ALICE DETECTOR}

\author{RACHID GUERNANE on behalf of the ALICE
  Collaboration\footnote{Presented at the $\mathrm{XL^{th}}$ Rencontres de
    Moriond on QCD and High Energy-Hadronic Interactions}}

\address{Laboratoire de Physique Corpusculaire de Clermont-Ferrand,\\
  Universit\'e Blaise Pascal and CNRS-IN2P3, 24 Avenue des Landais\\
  63177 AUBI\`ERE CEDEX FRANCE\\
  guernane@clermont.in2p3.fr}

\maketitle\abstracts{Heavy flavour pairs produced in hadronic
    reactions provide a valuable laboratory for the study of strong
    interactions. Due to their relatively large mass, the production
    of heavy quarks should be reliably calculable in the
    perturbative approach. Charm and beauty quarks once
    produced in a heavy ion collision have to propagate through the
    surrounding quark-gluon matter. Heavy quark states are then a
    sensitive probe of the properties of the dense medium. ALICE is a
    general-purpose experiment equipped to reconstruct, among other
    signals, leptons from open charm and beauty via their
    leptonic decays in p-p, \mbox{p-A} and \mbox{A-A} collisions. In
    these proceedings, we present
    feasibility studies for ALICE measurements of beauty production in
    central Pb-Pb collisions at $\sqrt{s_{NN}}=5.5\,\mathrm{TeV}$
    using semileptonic decays.}

\section{Why studying heavy quarks in high energy
  nuclear collisions?}
ALICE (A Large Ion Collider Experiment) is a detector dedicated
to the study of nucleus-nucleus interactions at the LHC. It will
investigate the physics of strongly interacting matter at extreme
energy densities, where the formation of a new phase of matter,
the QGP (Quark-Gluon Plasma) is expected. Heavy flavour physics is an
    important part of the ALICE experimental programme in many
    respects. As an intrinsically perturbative 
    phenomenon heavy quark production is a key benchmark for
    testing QCD and parton model concepts~\cite{Bloch,Onofrio,Wagner} in a novel kinematic region of large
    $Q^2$ and very low Bjorken-$x$ (as low as about $10^{-5}$)
    becoming accessible at the LHC. In the context of heavy ion  
    collisions, open heavy quark production is sensitive to gluon
    shadowing effects in
    nuclei~\cite{opencharm1,opencharm2}. Moreover, due to their long
    lifetime 
    heavy quarks live through the thermalization phase of a QGP
    and therefore carry information about the deconfined
    medium~\cite{pprchap1}. Measuring the beauty hadron production
    down to the very low $p_\mathrm{t}$ -- region where the ALICE detector has
    been optimized for tracking and particle identification
    (cf. section~\ref{alice}) -- is essential to minimize the
    extrapolation uncertainty on the total cross section. In addition, the low
    $p_\mathrm{t}$ region will be
    influenced by non-perturbative effects which can differ from
    p-p to A-A collisions. For heavy quarkonium physics, the measurement of the beauty quark yield in
    the same kinematic region as for the $\Upsilon$ measurement will
    provide a valuable reference against which to compare $\Upsilon$
    production in order to observe a possible suppression in A-A
    collisions. In the following, we will present the expected
    performance of the ALICE detector
    -- presently under construction -- for beauty
    production measurements in semileptonic decay channels.  

\section{Production of $\mathbbf{b}$-quarks at LHC}
Heavy quarks are produced in hadron-hadron collisions at LHC energies
by strong interaction processes described within the QCD
theory. Production cross section can be computed in the framework of the factorization
theorem of the QCD parton model as the convolution of
parton-parton scattering cross sections (calculable in pQCD) with
parton distribution functions. QCD higher order corrections to the Born reactions
have to be considered at LHC energies. Fixed-order NLO calculations of
heavy quark production cross sections are available~\cite{mnr} and
have been used to define the ALICE baseline~\cite{carrer}. In the analysis presented
hereafter, the \textsc{Pythia}~6.214 event generator~\cite{pythia} has been
used to produce heavy quarks after tuning to reproduce the relevant NLO
differential distributions (namely single quark $p_\mathrm{t}$,
rapidity, heavy quark pair invariant mass and azimuthal
correlation). Rates in heavy ion collisions are  
then obtained from binary scaling in accordance with the Glauber
multiple scattering model~\cite{glauber}. ``Known'' initial state effects
are included (namely primordial transverse momentum and nuclear
shadowing). The expected production rates are presented in
Table~\ref{tab:rates}. 

\begin{table*}[!htb]
  \renewcommand{\tabcolsep}{2pc} 
  \renewcommand{\arraystretch}{1.2} 
  \caption{Summary of heavy flavour production rates in the 5\,\% most central Pb-Pb collisions
    with an average luminosity of 
    \mbox{$\mathscr{L}_\mathrm{Pb\mbox{\scriptsize -}Pb}=5\times
    10^{26}\,\mathrm{cm^{-2}\cdot s^{-1}}$}. Minimum bias rates are
    also quoted for comparison. Pb-Pb geometric cross section is
    taken to be 7.39\,b.} 
  \begin{center}
    \scalebox{.85}{%
      \begin{tabular}{@{}|l|c|c|c|c|}
        \hline
        & \multicolumn{2}{c|}{$c\bar{c}$} &
        \multicolumn{2}{c|}{$b\bar{b}$} \\ \hline\hline
        N-N cross section [mb] & \multicolumn{2}{c|}{$6.64$} &
        \multicolumn{2}{c|}{$0.21$} \\
        EKS98 shadowing factor & \multicolumn{2}{c|}{$0.65$} &
    \multicolumn{2}{c|}{$0.86$} \\ \hline
        Centrality bin & $0\div 5\,\%$ & Min. Bias & $0\div 5\,\%$ &
        Min. Bias \\ \hline\hline
        $N_{Q\bar{Q}}$/Pb-Pb collision & 115.8 & 25.3 & 4.8 & 1.0 \\ 
        $N_{Q\bar{Q}}$/$\mathrm{10^6\,s}\ [\times 10^{10}]$ & $2.1$ & $9.3$ & $0.09$ & $0.4$ \\ \hline
      \end{tabular}}
  \end{center}
  \label{tab:rates}
\end{table*}

\section{Overview of the ALICE detector for lepton detection}\label{alice}

Both electrons and muons are measured in ALICE,
electrons in the central barrel and muons in a dedicated forward
spectrometer. Central detectors, covering mid-rapidity ($|\eta|\leq 0.9$) over the
full azimuth, are embedded inside the L3 solenoidal magnet providing
a magnetic field of \mbox{$\leq 0.5\,\mathrm{T}$}. Electrons are identified in
ALICE combining the TRD (Transition Radiation Detector) and TPC (Time
Projection Chamber) particle identification capabilities. TRD
provides an $e$/$\pi$ rejection power of 100 supplemented by the specific
electron energy loss in the TPC which allows to reach a combined pion
rejection factor of $10,000$. The ALICE
forward muon spectrometer has been designed to detect muons 
emitted in the pseudo-rapidity region \mbox{$-4\leq\eta\leq -2.5$}. The
forward muon spectrometer is made of a passive front absorber of 10
interaction lengths to absorb hadrons and photons from the interaction
vertex, a high-granularity tracking system of 10 cathod pad chambers,
a large dipole magnet creating a field of 0.7\,T (field integral of
$\mathrm{3\,T\cdot m}$), and a trigger system performing the selection of heavy
flavour decay muons by a transverse momentum cut-off made of four
resistive plate chambers. Muons penetrating the whole
spectrometer length are finally identified with a momentum resolution
of about 1\,\% and 90\,\% efficiency (for $p_\mathrm{t}>3\,\mathrm{GeV/c}$). 

\section{Beauty production measurements from semileptonic decays}
The copious beauty production at the LHC will be measured with the
ALICE detector using $b$-hadron semileptonic decays characterized by large
branching ratios. Leptons can indeed be produced both from 
direct decay $b\rightarrow\ell^-$ ($\mathrm{BR}\simeq 10\,\%$) and
cascade $b\rightarrow c\rightarrow\ell^+$ ($\mathrm{BR}\simeq
8\,\%$). After extracting the beauty signal in lepton data sets,
$b$-hadron production cross sections are assessed by unfolding
$b\rightarrow\ell$ decay.

\subsection*{Single inclusive $b$-quark production via muon detection}

\begin{figure}[!hbt]
  \hspace{2.5cm}\begin{minipage}[c]{.8\linewidth}
    \setlength\unitlength{1cm} 
    \begin{picture}(9,9.5)
      \put(0,5){\includegraphics[width=.8\linewidth,clip=]{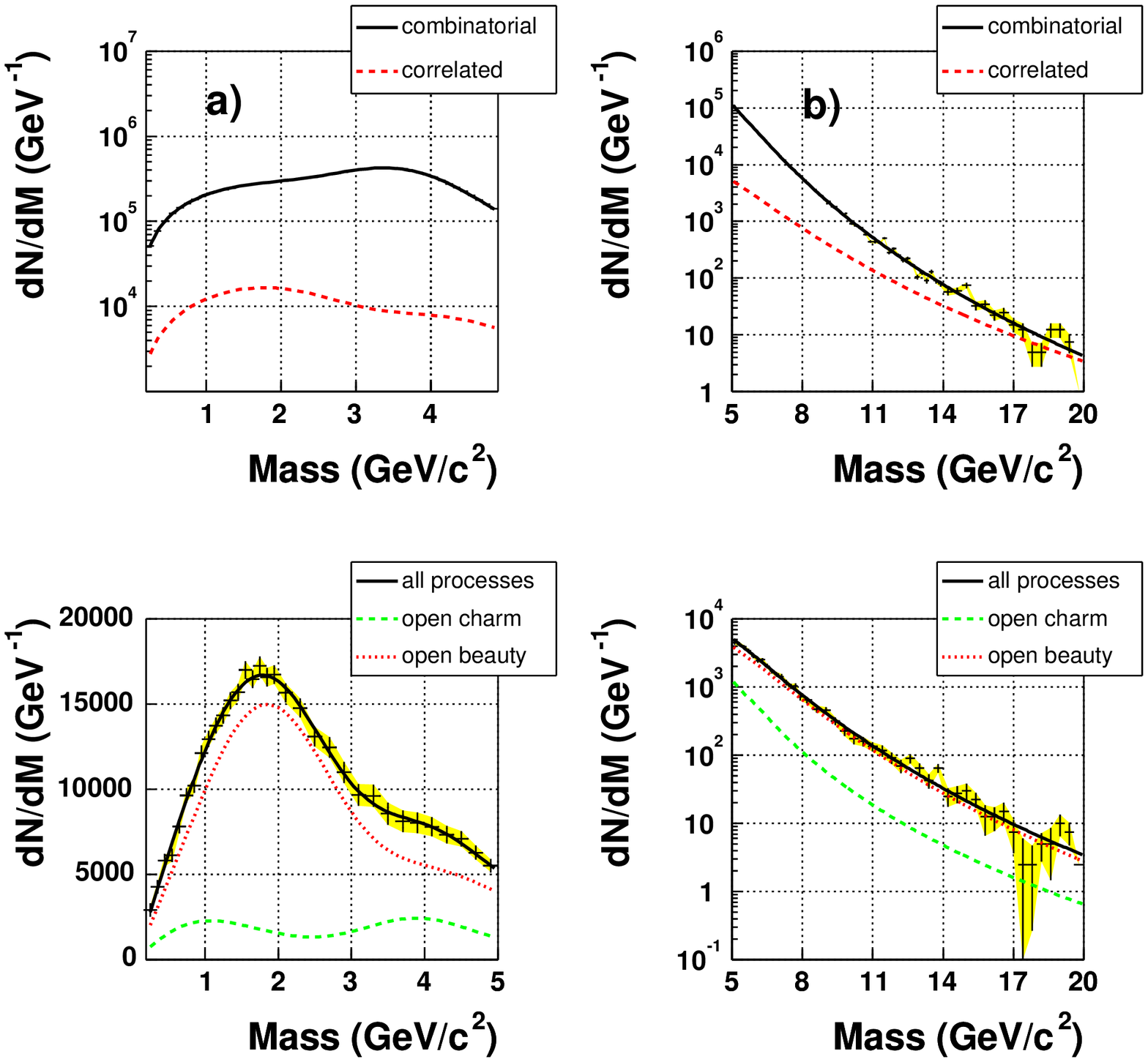}}
      \put(0,0){\includegraphics[width=.4\linewidth,clip=]{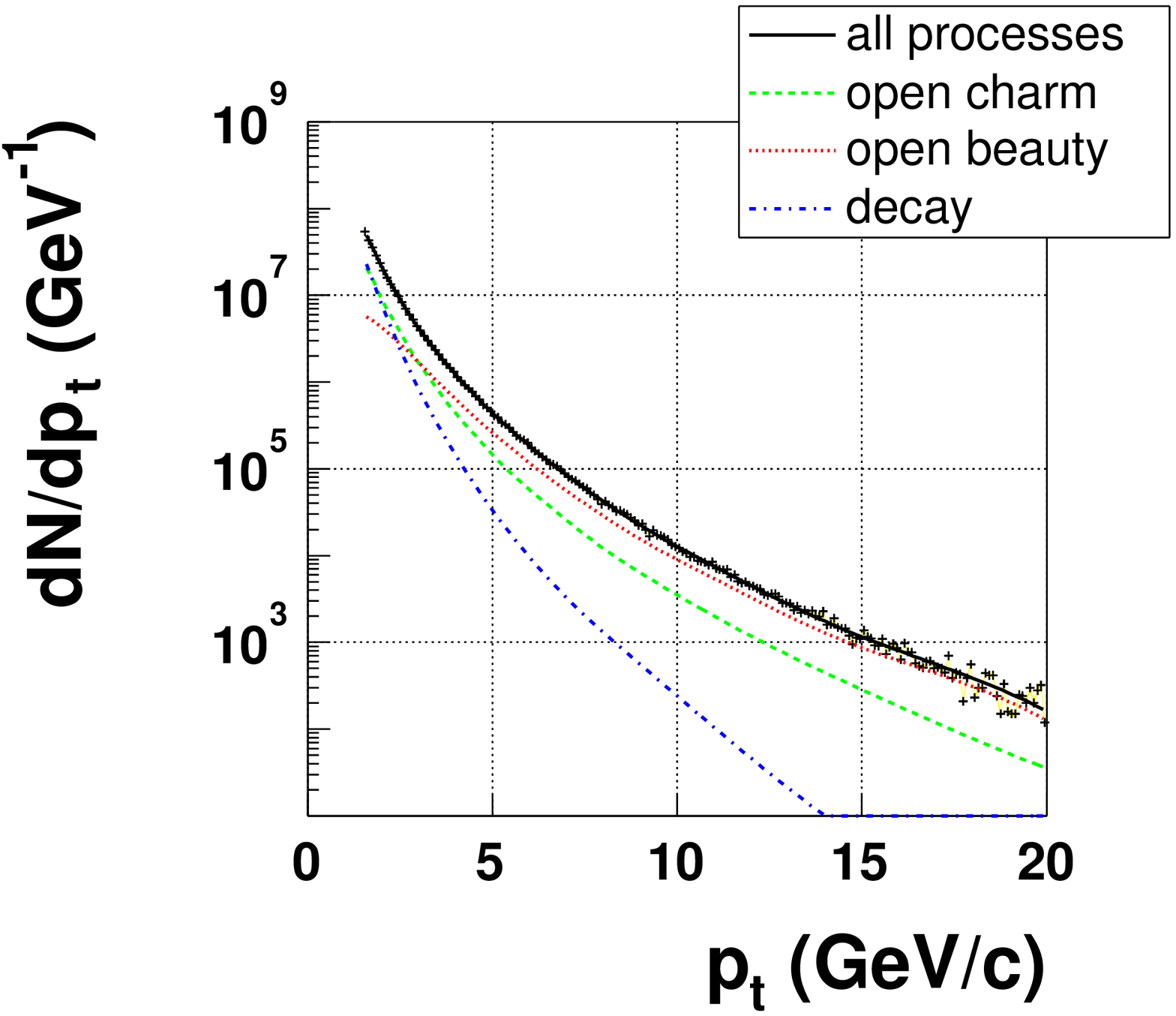}}
      \put(5.1,0){\includegraphics[width=.4\linewidth,clip=]{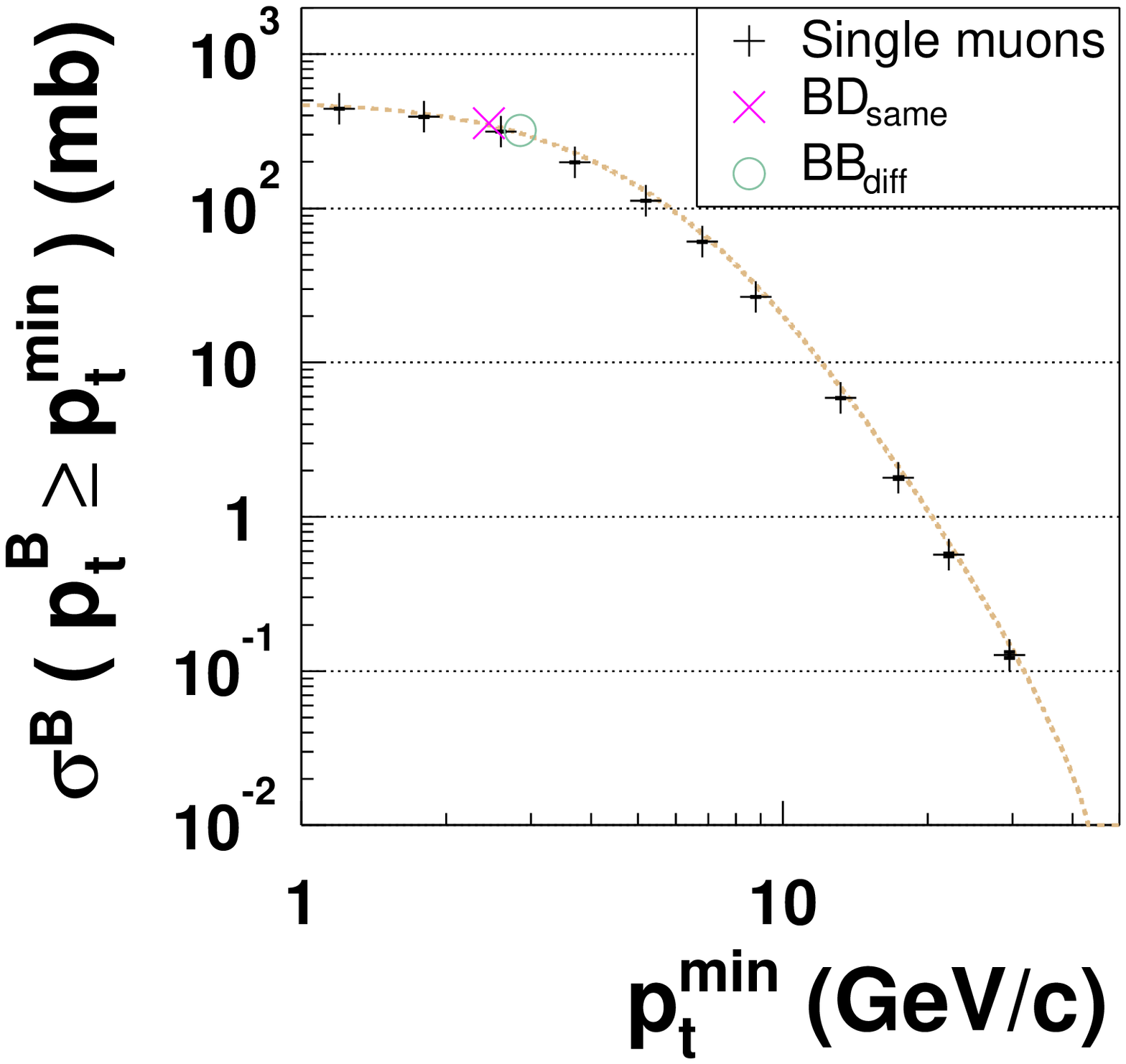}}
      \put(3.8,7.9){\usefont{T1}{phv}{bx}{n}{a)}}
      \put(8.9,7.9){\usefont{T1}{phv}{bx}{n}{b)}}
      \put(3.8,2.8){\usefont{T1}{phv}{bx}{n}{c)}}
      \put(8.9,2.8){\usefont{T1}{phv}{bx}{n}{d)}}
    \end{picture}
  \end{minipage}
  \caption{Background subtracted invariant mass distributions of
    $\mu^+\mu^-$ pairs produced in the 5\,\% most central Pb-Pb
    collisions at $5.5\,\mathrm{TeV}$ in the low (a) and high mass
    regions (b). A $\mathrm{p_t>1.5\,GeV/c}$ cut-off has been applied
    to muon tracks. Charm and beauty signals are plotted in green
    dashed and red dotted line respectively. (c) Single muon
    transverse momentum distribution. (d) Inclusive $b$-hadron cross
    section in $-4<y^B<-2.5$ as a function of
    $\mathrm{p^{min}_t}$. Also shown the \textsc{Pythia} prediction
    (dotted line) used to produce the signal.}
  \label{fig:xsec}
\end{figure}

Beauty production in Pb-Pb collisions will be measured in ALICE
using semileptonic decay muons in the pseudo-rapidity region
\mbox{$-4\leq\eta^\mu\leq -2.5$}. Both inclusive muon and opposite 
sign dimuon productions are considered, the dimuon sample being divided
into two topologically distinct contributions: $b$-chain decays (named
$\mathrm{BD_{same}}$ hereafter) of low
mass \mbox{$M_{\mu\mu}<5\,\mathrm{GeV/c^2}$} and high transverse
momentum (cf. Fig.~\ref{fig:xsec}(a)) and muon pairs where the two muons originate from
different quarks ($\mathrm{BB_{diff}}$) emitted at large angles
resulting in large invariant masses
\mbox{$M_{\mu\mu}>5\,\mathrm{GeV/c^2}$}
(cf. Fig.~\ref{fig:xsec}(b)). Beauty signal is enhanced with respect
to other sources (charm and $\pi/K$ decay-in-flight) applying a low
$p_\mathrm{t}$ cut-off of $1.5\,\mathrm{GeV/c}$~\cite{rach}. Using
Monte Carlo predicted line shapes for $b\bar{b}$, $c\bar{c}$, and decay 
background, fits are carried out to find the $b\bar{b}$ fraction in the different
data sets as presented in Fig.~\ref{fig:xsec}(a), (b), and
(c). Finally muon level cross sections are converted into inclusive
$b$-hadron cross section following the method initially developped for
the UA1 experiment~\cite{ua1}. The $b$-hadron cross section
measurement is plotted in Fig.~\ref{fig:xsec}(d), it shows no
systematic bias with respect to the initial distribution used in the
simulation to produce beauty signal even if a detailed study of systematic
uncertainties has still to be done. $b$-decay muon statistics is
large over the whole $p_\mathrm{t}$ range allowing a 
tight mapping of the production cross section up to
$p_\mathrm{t}=30\,\mathrm{GeV/c}$. 

\subsection*{Single inclusive $b$-quark production via electron detection}
\begin{figure}[!hbt]
  \begin{minipage}[c]{.8\linewidth} 
    \setlength\unitlength{1cm} 
    \begin{picture}(9,5.3)
      \put(2,0){\includegraphics[width=.9\linewidth]{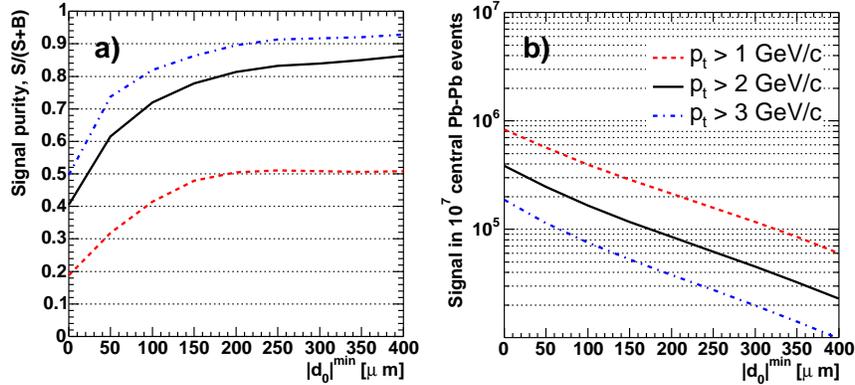}}
      \put(3.3,4.4){\usefont{T1}{phv}{bx}{n}{a)}}
      \put(9,4.4){\usefont{T1}{phv}{bx}{n}{b)}}
    \end{picture}
  \end{minipage}
  \caption{Signal to background ratio (a) and signal (b) for beauty
    decay electrons in $10^7$ central Pb-Pb events (5\,\%),
    corresponding to about one month of LHC data taking.}   
  \label{fig:cuts}
\end{figure}

Electrons from semileptonic decay of $b$-quarks are characterized by a
hard $p_\mathrm{t}$ spectrum and a large average impact
parameter\footnote{Defined as the distance of closest approach of
  the electron track to the primary vertex in the transverse
  plane. $d_0$ is measured by the ALICE Inner Tracking System (ITS)
  with a resolution $\lesssim 70\,\mu\mathrm{m}$ at $p_\mathrm{t}=1\,\mathrm{GeV/c}$.}
$\langle d_0\rangle$ ($\simeq 300\,\mu\mathrm{m}$) with respect to
other electron sources (pion misidentification, charm, light
mesons, and photon conversions). The beauty signal purity and statistics as a
function of the impact parameter cut-off for different values of
the $p_\mathrm{t}$ threshold are shown in Fig.~\ref{fig:cuts} a)
and b) respectively. $p_\mathrm{t}>2\,\mathrm{GeV/c}$ and
$200<|d_0|<600\,\mu\mathrm{m}$ provide an electron sample of $8\times
10^4$ with a 90\,\% purity~\cite{enote}. An upper limit on $d_0$ is needed to reduce
long lived strange particles and tracks suffering from large angle
scatterings in detector materials. 

\section{Conclusion}

The ALICE detector has good capabilities for heavy flavour
measurements~\cite{ppr2} with the unique ability to address the $b$-quark sector in
p-p, p-A and A-A collisions with relatively low transverse momentum
thresholds. The measurement of $b$ production provides an
important test of the theory of QCD in heavy ion collisions where new
effects are expected as compared to nucleon-nucleon
interactions. A precise measurement of the inclusive
$\mathrm{d\,\sigma}^B/\mathrm{d\,p_t}$ cross section will allow to
probe in-medium quenching effects. Channels discussed in these
proceedings will be supplemented by measuring dilepton correlations,
multi-muon topologies, and $b$-tagged jets. Quantitative studies are
currently underway.


\section*{Acknowledgments}
The author would like to thank F.~Antinori and
A.~Dainese, members of the ALICE Collaboration, for providing me the
relevant material to prepare this talk and valuable discussions.


\section*{References}
\begin{thebibliography}{99}

\bibitem{Bloch} I.~Bloch, These Proceedings.
\bibitem{Onofrio} M.~D'Onofrio, These Proceedings.
\bibitem{Wagner} J.~Wagner, These Proceedings.
\bibitem{opencharm1} Z.~Lin and M.~Gyulassy, Phys. Rev. \textbf{C~51} (1995) 2177~;
  Phys. Rev. \textbf{C~52} (1995) 440.
\bibitem{opencharm2} Z.~Lin and M.~Gyulassy, Phys. Rev. Lett. \textbf{77} (1996) 1222.
\bibitem{pprchap1} ALICE Collaboration 2004 \emph{J. Phys. G}
  \textbf{30} 1517-1763. 
\bibitem{mnr} M.L.~Mangano, P.~Nason and G.~Ridolfi, Nucl. Phys. \textbf{B~373} (1992) 295.
\bibitem{carrer} N.~Carrer and A.~Dainese, ALICE-INT-2003-019, \texttt{hep-ph/0311225}.
\bibitem{pythia} T.~Sj\"ostrand, P.~Ed\'en, C.~Friberg, L.~L\"onnblad,
  G.~Miu, S.~Mrenna and E.~Norrbin, Computer Physics Commun. {\bf 135} (2001) 238.
\bibitem{glauber} R.J.~Glauber, Lecture in Theoretical Physics,
    Vol.~1, eds W.E.~Brittin and L.G.~Dunham (Interscience, New-York,
    1959) p.~315.
\bibitem{rach} R.~Guernane \emph{et al.}, ALICE Internal Note, \emph{in preparation}. 
\bibitem{ua1} UA1 Collaboration, C. Albajar \emph{et al.}, Phys. Lett. \textbf{B~213} (1988) 256, \\
  UA1 Collaboration, C. Albajar \emph{et al.}, Phys. Lett. \textbf{B~256} (1991) 121. 
\bibitem{enote} F.~Antinori \emph{et al.}, ALICE Internal Note,
  \emph{in preparation}.
\bibitem{ppr2} ALICE Physics Performance Report Vol.~II, ALICE
  Collaboration, \emph{in preparation}.
\end{thebibliography}
\end{document}